\documentclass[conference]{IEEEtran}
\usepackage{cite}
\usepackage{amsmath,amssymb,amsfonts}
\usepackage{algorithmic}
\usepackage{graphicx}
\usepackage{textcomp}
\usepackage{xcolor}
\usepackage{comment}
\usepackage{color,soul}

\usepackage[linesnumbered,ruled,vlined]{algorithm2e}
\SetCommentSty{mycommfont} % algorithm2e
\SetKwInput{KwInput}{Input} % algorithm2e
\SetKwInput{KwOutput}{Output} % algorithm2e

\def\BibTeX{{\rm B\kern-.05em{\sc i\kern-.025em b}\kern-.08em
    T\kern-.1667em\lower.7ex\hbox{E}\kern-.125emX}}

\begin{document}
\title{Distributed Load Orchestration for Vision Computing in Multi-Access Edge Computing}

\author{
    \IEEEauthorblockN{Ricardo N. Boing}
    \IEEEauthorblockA{
        \textit{UFSC} \\
        Florianópolis, Brazil \\
        ricardoboing.ufsc@gmail.com
    }
    \and
    \IEEEauthorblockN{Hugo Vaz Sampaio}
    \IEEEauthorblockA{
        \textit{UFSC} \\
        Florianópolis, Brazil \\
        hvazsampaio@gmail.com
    }
    \and
    \IEEEauthorblockN{Fernando Koch}
    \IEEEauthorblockA{
        \textit{IBM} \\
        USA \\
        fkoch@acm.org
    }
    \and
    \IEEEauthorblockN{René N. S. Cruz}
    \IEEEauthorblockA{
        \textit{UFSC} \\
        Florianópolis, Brazil \\
        rnoliosc@gmail.com
    }
    \and
    \IEEEauthorblockN{Carlos B. Westphall}
    \IEEEauthorblockA{
        \textit{UFSC} \\
        Florianópolis, Brazil \\
        carlos.westphall@ufsc.br
    }
}

\maketitle

\begin{abstract}

Multi-access Edge Computing (MEC) is a type of network architecture that provides cloud computing capabilities at the edge of the network. We consider the use case of video surveillance for an university campus running on a 5G-MEC environment. A key issue is the eventual overloading of computing resources on the MEC nodes during peak demand. We propose a new strategy for distributed orchestration in MEC environments based on how load balancing strategies organize processing queue. Then, we elaborated a strategy for deadline-aware queueing prioritization that organizes requests based on pre-established thresholds. We introduce a simulation-based experimentation environment and conduct a number of tests  demonstrating the benefit of our approach by reducing the number of referrals and improving the effectiveness in meeting deadlines.
\end{abstract}

\section{Introduction}
\label{sec:introduction}

This research is motivated by a social issue: the need to improve safety and monitoring in our  university campus through video surveillance. We architect our solution by using 5G network and Multi-access edge computing (MEC) \cite{etsi2018mec, etsi2019multi} as the enabler technologies. This environment provides the benefits of low-latency, availability, local processing, experimental structure, and a great learning opportunity. On top of this structure, we aim to explore solutions to improve situation awareness, incident management, and event automation \cite{2019_sultana_wahid}, as well as use cases around e.g. people counting and queue detection for the university restaurant, activity visualizer for bus stations \cite{2016_neethu_athi_bijlani}, perimeter and intrusion detection for secured areas, and others.

Nonetheless, MEC nodes have constrained computing resources that can become overloaded during peak times\cite{2016_li_wu_zhao}. The common approach for load orchestration is to apply strategies around an intermediate node to receive the load and transfer it to the \textit{more available} load by some metric \cite{2021_liu_mao_xu_hu_chen, 2021_kaur_sandhu_mohana, 2022_tahmasebi_sarram_mostafavi}. These approaches are less suitable in scenarios that require very-low latency \cite{2020_beraldi_canali_lancellotti_mattia_elsevier, 2022_etsi_mec}. Other approaches attribute to the MEC the decision to distribute or not the load \cite{2020_beraldi_canali_lancellotti_mattia_acm}. Nonetheless, these strategies do not consider the prioritization of requests, eventually inflicting Service Level Agreement (SLA) around pre-established deadlines and maximum number of referrals.

Hence, there is a need to explore new strategies for distributed load orchestration that are  deadline-aware, that is able maximizes the number of requests served within a SLA-established  deadline and reduce the number of forwards to neighboring nodes. We propose to modify the \emph{Sequential Forwarding Algorithm} \cite{2020_beraldi_canali_lancellotti_mattia_ieee}, which originally uses a FIFO-type request queue. Instead, we implemented a strategy for preferential queueing that organizes requests based on pre-established thresholds.

This study provides the following contributions to the state-of-the-art:
\begin{itemize}
\item presents a novel strategy for deadline-aware distributed load orchestration for Video Computing on 5G-MEC environments;
\item introduces a simulation-based experimentation environment, and;
\item analyzes the results of multiple test configurations to compare the gain in performance around response time and SLA compliance against existing approaches. 
\end{itemize}

This paper is organized as follows. Section \ref{sec:background} provides a review of the existing state-of-the-art and introduces the Sequential Forwarding Algorithm. We present our proposed variation on Section \ref{sec:proposed} and the experimental environment in Section \ref{sec:material_and_methods}. Section \ref{sec:result} elaborates on the results of our tests and benefits of our approach. Our conclusions and proposal for future work are discussed on Section \ref{sec:conclusion}.

\section{Background and Related Work}
\label{sec:background}

Multi-access edge computing (MEC) is an evolution in cloud computing that uses mobility, cloud services, and edge computing to move application hosts away from a centralized datacenter to the edge of the network  \cite{2022_etsi_mec, 2017_mao_you_zhang_huang_letaief}. The MEC architecture provides the mechanisms to  activate, update, and deactivate MEC applications and configuration rules required to regulate the traffic between applications, intra-node communication, and loading of application on nodes. This allows IP traffic routing, distributed load orchestration, or tapping to the MEC applications or to locally accessible networks. 

A common challenge in this sort of environment is on how to handle fluctuating workload demands. This issue calls for strategies of distribute load orchestration able to offload outbound demands to nearby nodes and/or Cloud-based services in order to ensure SLA-compliance and sustain Quality-of-Service \cite{2016_li_wu_zhao}. The process of intra-node load orchestration relates to that of Grid Computing \cite{assuncao2004grids} and other peer-to-peer approaches, which served as inspiration for this research. We focus on how to optimize processing through an innovative approach for the prioritization queue.

We carried out a review of the prior-art and identified the following patterns of load orchestration strategies:

\begin{enumerate}

\item \emph{Centralizing load orchestration node}~\cite{2019_mostafa_nour, 2021_herbert_raj}: we concluded that this approach is not viable for Video Computing in 5G-MEC because the implying a bottleneck and extra load on the communication network  could result in delays that will affect the response time of requests during peak times. 

\item \emph{Distributed load orchestration nodes}~\cite{2021_liu_mao_xu_hu_chen, 2021_kaur_sandhu_mohana}, where the local brokers decide on load orchestration for their location and know the load of brokers at other locations. Similarly, this approach is inappropriate for our use case as, despite reducing the delay in response time ,it continues to generate additional traffic on the network since the broker is closer to the nodes that will receive the load.

\item \emph{Criteria-based load orchestration}~\cite{2022_tahmasebi_sarram_mostafavi}, where MEC nodes will forward requests only when a certain criterion is reached, without an intermediate node. The approach considers the node's ability to process a request within the deadline; if it fails, the node will forward the request to a neighboring node. Without knowing the availability of its neighboring nodes, the MEC node uses machine learning to select the neighbors where the load will be forwarded. The drawback of this approach is that it does not consider the organization of the request queue and, eventually  inflicting Service Level Agreement (SLA) around pre-established deadlines.

\end{enumerate}

The work in \cite{2017_gasior_seredynski} proposes a method for distributing requisitions to maximize compliance with deadlines. The approach applies a queue organized by the deadline of the requests. The drawback is to generate intensive offloading whilst searching for load orchestration between the nodes, thus inflicting high network traffic.

The \emph{Sequential Forwarding Algorithm}, introduced in \cite{2020_beraldi_canali_lancellotti_mattia_ieee}, replaces centralized load orchestration decision with the individual decision of each node. The objectives is to reduce traffic generated on the network and minimize waiting time. The proposed algorithm considers that Fog nodes must individually receive user requests and insert them into a FIFO queue. If the node is sufficiently loaded to miss the request deadline, the request is not added to the queue, and forwarding is performed to a randomly chosen neighbor node. Each request can be forwarded a maximum $M$ number of times so that if the $M$ number is reached, the last node to receive it cannot make a new forwarding and is forced to process it.

The work in \cite{2020_beraldi_canali_lancellotti_mattia_elsevier} proposes that if the number $M$ of referrals is reached, and the node $M$ predicts that it will not be able to process the request within the deadline, the request should be discarded.

The work in \cite{2020_beraldi_canali_lancellotti_mattia_acm} introduces a modified algorithm to select a set of neighboring nodes. The selected neighbors will be consulted, and the one with the lowest load will be chosen to receive the load. If there is no suitable neighbor to receive the request, it is discarded.

Hence, we concluded for the need to explore innovative strategies for distributed load orchestration through the optimization of the \emph{prioritization processing queue} in \emph{Local Load Orchestrators}. We seek to create an approach that is deadline-aware, that is able maximizes the number of requests served within a SLA-established  deadline and reduce the number of forwards to neighboring nodes. We introduce our approach around a modification of common distribution algorithm applied to load orchestration.
\section{Proposal}
\label{sec:proposed}
We introduce a modification of the \emph{Sequential Forwarding Algorithm} to reduce the number of forwardings and maximize the number of requests served within the deadline. We apply the first version of the algorithm, which does not discard requests with expired deadlines. We update the algorithm by replacing the request queue, which was originally a FIFO type, with a preferential queue. That is, new requests with shorter deadlines can be allocated in front of requests already allocated if the deadline of the others is not affected.

\begin{figure}[!ht]
    \centering
    \includegraphics[width=.49\textwidth]{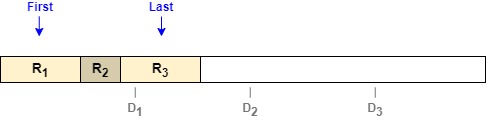}
    \caption{Escalation algorithm}
    \label{fig:queue1}
\end{figure}

Figure \ref{fig:queue1} illustrates the proposed approach. The first request to be processed is the leftmost one, while the last request is allocated on the right. Each request $R_i$ has a response time $D_i$ and points to the next request. Depending on the size of the established deadline $D$ and the processing time, it is possible to allocate new requests in front of others without losing deadlines.

\begin{figure}[!ht]
    \centering
    \includegraphics[width=.49\textwidth]{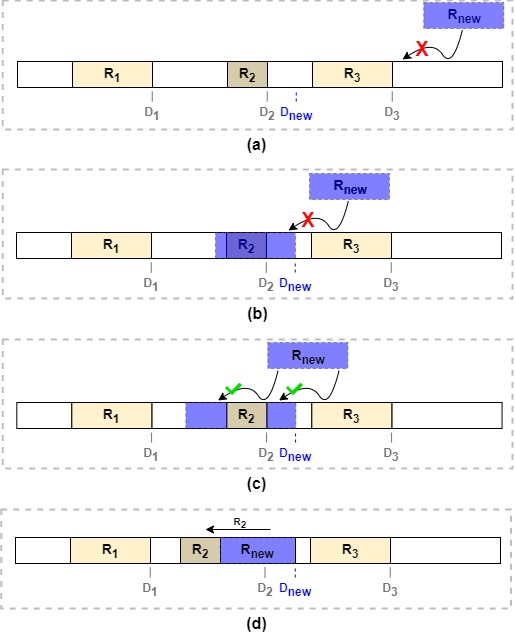}
        \caption{Queue processing strategy}
    \label{fig:queue2}
\end{figure}

Figure \ref{fig:queue2} depicts the queue processing mechanism as follows: an attempt is made to allocate a new request $R_{new}$ at the end of the queue (Figure \ref{fig:queue2}a);  the allocator verifies that the term of $R_{new}$ would be extrapolated if the allocation was made after $R_3$; then, the allocator checks that there is a time gap between the end of processing $R_2$ and the beginning of $R_3$ that could be used to allocate $R_{new}$ (Fig. \ref{fig:queue2}b); however, it infers that the time-space is insufficient to perform the allocation.

Next, the allocator considers the time-space between $R_2$ and $R_3$ and keeps looking for more time-spaces (Fig. \ref{fig:queue2}c).  A time-space between $R_1$ and $R_2$ is found and is longer than necessary. Then, the allocator allocates $R_{new}$ between $R_2$ and $R_3$ and the available spaces between $R_1$ and $R_2$, as well as the spaces between $R_2$ and $R_3$, are reduced according to the processing time and the term of $R_{new}$ (Fig. \ref{fig:queue2}d). The deadline $D_2$ did not follow the block $R_2$ as illustrated in the previous steps. That is, the deadline for responding to $R_2$ remains the same, but there is not enough space before $R_2$ to allocate new requests that make $R_2$ finish executing within the deadline $D_2$.

\begin{figure}[!ht]
    \centering
    \includegraphics[width=.49\textwidth]{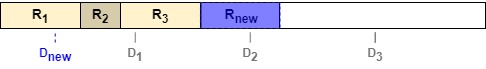}
    \caption{Processing queue in the worst case scenario}
    \label{fig:queue3}
\end{figure}

Figure \ref{fig:queue3} illustrates the queue in the worst-case scenario, when the allocation algorithm identifies that $R_{new}$ cannot be allocated at the end of the queue and that there are not enough time slots available. To exemplify, we created a case with very short deadline $D_{new}$ so that $R_{new}$ needed to be processed before $R_1$. There is not enough space to allocate $R_{new}$, so there are two alternatives. The first alternative is to forward $R_{new}$ to a randomly chosen neighbor node being necessary that the maximum number of forwards has not been reached. If the maximum number of forwards has been reached, then $R_{new}$ must be allocated at the end of the queue, and all available time slots will be removed. This decision will result in losing the $R_{new}$ deadline, but the request will be processed and answered even with the missed deadline. In addition, as there was no reallocation, the deadline for the other requests will continue to be respected.

The algorithm for preferential queueing is presented in Algorithm \ref{alg:push_request}. It receives as parameters the request to be allocated, the time when the CPU will be available to process the next request, and whether the request should be added to the queue even in case of non-compliance with the deadline.

\begin{algorithm}[!ht]
\caption{push\_request(request, cpuFreeTime, forcedPush)}
\label{alg:push_request}
newBlock = RequestBlock(request)

leftBlock = this.lastBlock

rightBlock = null

spaceNeeded = newBlock.get\_size()

hasRightSpace = false

status = search\_alloc\_space(leftBlock, newBlock, rightBlock, spaceNeeded, hasRightSpace, cpuFreeTime, forcedPush)

\If{$status == true$} {
	return true
}

\If{$forcedPush == false$} {
	return false
}

\eIf{$is\_empty() == true$} {
	start = cpuFreeTime
}{
	start = leftBlock.get\_end()
}

end = start + spaceNeeded

newBlock.set\_end(end)

alloc\_request(leftBlock, newBlock, rightBlock)

return true

\end{algorithm}

\begin{algorithm}[!ht]
\caption{search\_alloc\_space(leftBlock, newBlock, rightBlock, spaceNeeded, hasRightSpace, cpuFreeTime, forcedPush)}

usefulArea = get\_useful\_area(leftBlock, newBlock, rightBlock, cpuFreeTime)

end = usefulArea.get\_end()

freeSpace = usefulArea.get\_size()

\If{$freeSpace \ge spaceNeeded$} {
	shift\_or\_alloc(leftBlock, newBlock, rightBlock, end, spaceNeeded, hasRightSpace)
	
	return true
}

\If{$leftBlock == null$} {
	\If{$forcedPush == true$ and $rightBlock \ne null$} {
		shiftValue = rightBlock.get\_start() - cpuFreeTime
		
		end = rightBlock.get\_end() - shiftValue
	    
	    rightBlock.set\_end(end)
	}
	
	return false
}

\eIf{$freeSpace > 0$} {
	\_hasRightSpace = true
}{
	\_hasRightSpace = hasRightSpace
}

\_freeNeeded = spaceNeeded - freeSpace

\_leftBlock = leftBlock.get\_left\_block()

\_rightBlock = leftBlock

status = search\_alloc\_space(\_leftBlock, newBlock, \_rightBlock, \_freeNeeded, \_hasRightSpace, cpuFreeTime, forcedPush)

\If{$status == true$} {
	\If{$forcedPush == true$ and $rightBlock \ne null$} {
		shiftValue = rightBlock.get\_start() - leftBlock.get\_end()
		
		end = rightBlock.get\_end() - shiftValue
    	
    	rightBlock.set\_end(end)
	}
	
	return false
}

shift\_or\_alloc(leftBlock, newBlock, rightBlock, end, spaceNeeded, hasRightSpace)

return true

\end{algorithm}

\begin{algorithm}[!ht]
\caption{get\_useful\_area(leftBlock, newBlock, rightBlock, cpuFreeTime)}

\eIf{$leftBlock \ne null$} {
	start = leftBlock.get\_end()
}{
	start = cpuFreeTime
}

\eIf{$rightBlock \ne null$} {
	end = rightBlock.get\_start()
}{
	end = math.inf
}

end = min(end, newBlock.get\_end())

\If{$start > end$} {
	start = 0
	
	end = 0
}

width = end - start

return Block(width, end)

\end{algorithm}

\begin{algorithm}[!ht]
\caption{shift\_or\_alloc(leftBlock, newBlock, rightBlock, end, spaceNeeded, hasRightSpace)}

\eIf{$hasRightSpace == true$} {
	end = rightBlock.get\_end() - spaceNeeded
	
	rightBlock.set\_end(end)
}{
	\If{newBlock.get\_right\_block() == null and newBlock.get\_left\_block() == null} {
		newBlock.set\_end(end)
		
		alloc\_request(leftBlock, newBlock, rightBlock)
	}
}
\end{algorithm}

\begin{algorithm}[!ht]
\caption{alloc\_request(leftBlock, newBlock, rightBlock)}

\eIf{$leftBlock \ne null$} {
	leftBlock.set\_right\_block(newBlock)
}{
	this.firstBlock = newBlock
}
\eIf{$rightBlock \ne null$} {
	rightBlock.set\_left\_block(newBlock)
}{
	this.lastBlock = newBlock
}

newBlock.set\_left\_block(leftBlock)

newBlock.set\_right\_block(rightBlock)

\end{algorithm}
\section{Experiments}
\label{sec:material_and_methods}

\begin{figure}[!ht]
    \centering
    \includegraphics[width=.49\textwidth]{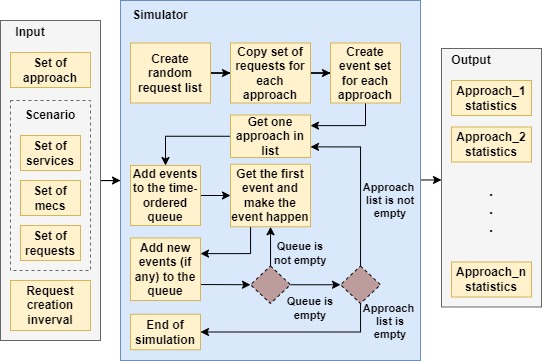}
    \caption{MEC-LB Simulator.}
    \label{fig:simulator}
\end{figure}

We developed the \emph{MEC-LB Simulator} to provide the experimentation framework for our exploration (see Figure \ref{fig:simulator}). All scenarios contain a set of MEC nodes that provide the same services. It works by imitating the following scenario: 

\begin{itemize}
\item users send the requests to the nearest MEC;
\item an application running on this node receives and processes the requests interactively; 
\item the service has a maximum processing time and a deadline for returning a response
\item we neglect delays generated by the network, scheduling, and allocation of requests; 
\item we consider all MEC nodes have equivalent computing resources. 
\end{itemize}

Each service has a maximum processing time and a deadline for returning a response. Before starting the simulations, a list of requests each MEC node will receive during the simulation is generated. A copy of the requisition list simulates each load distribution approach, ensuring similar conditions for comparative purposes between the approaches. However, for forwarding requests, the MEC node that will receive the forwarding is chosen randomly at the time the forwarding takes place. In this initial study, we also consider that all services must reach the worst case in the processing time.

To simulate the Video Surveillance use case, we consider that the camera system will use devices capable of capturing images with different resolutions, such as HD, Full HD, and 4K. We assume that the size of images will directly affect GPU usage, so the maximum time to process a 4K image is longer than the maximum time to process an HD or Full HD image.

Table \ref{table:service_list} present the  processing times and response times for each service are listed, which were named $S_1$, $S_2$, $S_3$, $S_4$, $S_5$ and $S_6$. The values for the processing time are hypothetical and proportional to the number of pixels of each resolution. The processing time and the established deadlines are measured in a generic time scale that we call $UT$ (unit of time).

\begin{table}[!ht]
\centering
\caption{Service data}
\begin{tabular}{c|c|c|c|c|}
\cline{2-5}
 & \textbf{Number of pixels} & \textbf{Environment} & \textbf{Process time} & \textbf{Deadline} \\ \hline
\multicolumn{1}{|c|}{\textbf{S1}} & 8,294,400 & Busy     & 180 & 9,000 \\
\multicolumn{1}{|c|}{\textbf{S2}} & 2,073,600 & Busy     & 44  & 9,000 \\
\multicolumn{1}{|c|}{\textbf{S3}} & 921,600   & Busy     & 20  & 9,000 \\
\multicolumn{1}{|c|}{\textbf{S4}} & 8,294,400 & Isolated & 180 & 4,000 \\
\multicolumn{1}{|c|}{\textbf{S5}} & 2,073,600 & Isolated & 44  & 4,000 \\
\multicolumn{1}{|c|}{\textbf{S6}} & 921,600   & Isolated & 20  & 4,000 \\ \hline
\end{tabular}
\label{table:service_list}
\end{table}

We created three variations of the experimentation scenarios. In scenarios 1 and 2 we consider the existence of three MEC nodes, named $M_1$, $M_2$ and $M_3$. In scenario 3, 3 more MEC nodes were added, named $M_4$, $M_5$ and $M_6$.  Table \ref{table:request_list} presents the numbers of requests made to each of the MEC nodes, referring to each of the services provided, are displayed. We executed an average of 40 simulations per experimentation environment and collected the data on (i) the rate of answered requests within an established deadline and (ii) the rate of referrals made by each MEC nodes. 

\begin{table}[!ht]
\centering
\caption{Number of requests for each service made to each mec}
\begin{tabular}{cc|c|c|c|c|c|c|}
\cline{3-8}
                                          &             & \textbf{S1} & \textbf{S2} & \textbf{S3} & \textbf{S4} & \textbf{S5} & \textbf{S6} \\ \hline
\multicolumn{1}{|c|}{}                    & \textbf{M1} & 500         & 300         & 200         & 500         & 300         & 200         \\
\multicolumn{1}{|c|}{\textbf{Scenario 1}} & \textbf{M2} & 200         & 300         & 500         & 200         & 300         & 500         \\
\multicolumn{1}{|c|}{}                    & \textbf{M3} & 300         & 500         & 200         & 300         & 500         & 200         \\ \hline
\multicolumn{1}{|c|}{}                    & \textbf{M1} & 250         & 300         & 700         & 250         & 300         & 700         \\
\multicolumn{1}{|c|}{\textbf{Scenario 2}} & \textbf{M2} & 100         & 300         & 1,000        & 100         & 300         & 1,000        \\
\multicolumn{1}{|c|}{}                    & \textbf{M3} & 150         & 500         & 700         & 150         & 500         & 700         \\ \hline
\multicolumn{1}{|c|}{}                    & \textbf{M1} & 250         & 300         & 700         & 250         & 300         & 700         \\
\multicolumn{1}{|c|}{}                    & \textbf{M2} & 100         & 300         & 1,000        & 100         & 300         & 1,000        \\
\multicolumn{1}{|c|}{\textbf{Scenario 3}} & \textbf{M3} & 150         & 500         & 700         & 150         & 500         & 700         \\
\multicolumn{1}{|c|}{}                    & \textbf{M4} & 100         & 100         & 100         & 100         & 100         & 100         \\
\multicolumn{1}{|c|}{}                    & \textbf{M5} & 100         & 100         & 100         & 100         & 100         & 100         \\
\multicolumn{1}{|c|}{}                    & \textbf{M6} & 100         & 100         & 100         & 100         & 100         & 100         \\ \hline
\end{tabular}
\label{table:request_list}
\end{table}

\section{Results}
\label{sec:result}

\begin{figure}[!ht]
    \centering
    \includegraphics[width=.49\textwidth]{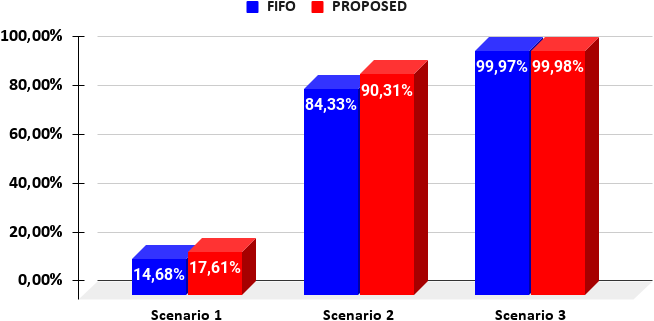}
    \caption{FIFO and preferential queue requests answered within deadline}
    \label{fig:success_rate}
\end{figure}

The results were obtained from an average of 40 simulations performed for each experiment scenario. The rates of requests fulfilled within the deadline when using the FIFO queue and the proposed face-to-face queue are shown in Figure \ref{fig:success_rate}. We considered that scenarios 1, 2, and 3 had 6000 requests, 8000 requests, and 9800 requests, respectively, to calculate the percentages. These sums can be obtained from the number of requests presented in Table \ref{table:request_list}. In every scenarios, the preferred queue proved superior to the FIFO queue as depicted in Figure \ref{fig:success_rate}.

In scenario 1, the difference was 2.92\% more deadlines met when the proposed queue was used. To evaluate the effect of the type of service requested by users, in scenario 2, we increased the number of requests for services $S_3$ and $S_6$, which require less processing time and reduced the number of requests for services $S_1$ and $S_4$, which require more processing time. There was a greater difference in the success rate, equivalent to 5.97\% more deadlines met when the proposed queue was used.

In scenario 3, we decided to keep the number of requests from scenario 2 made to each MEC node. The difference is that we doubled the number of MEC nodes from 3 to 6, and the 3 new MEC nodes received much lower requests for each type of service. It is possible to observe that the success rate was very similar in both queues; however, the proposed queue reached about 0.01\% more deadlines met.

\begin{figure}[!ht]
    \centering
    \includegraphics[width=.49\textwidth]{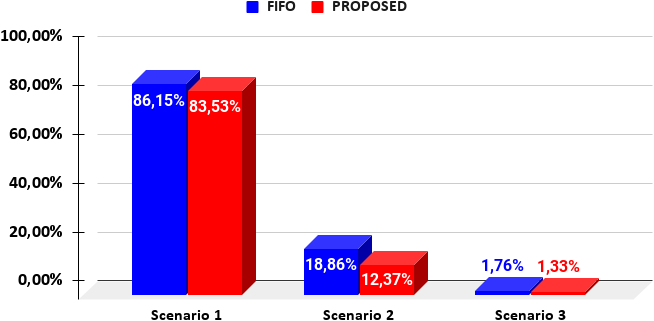}
    \caption{FIFO and preferential queue forwarding requests.}
    \label{fig:forward_rate}
\end{figure}

Figure \ref{fig:forward_rate} presents the forwarding rates performed in each experimentation scenario. For the experiments, it was considered that a maximum of 2 forwarding requests would be possible. Therefore, based on the number of requests made in each scenario, the maximum number of possible referrals for scenarios 1, 2, and 3 is 12000, 16000, and 19600, respectively.

The preferred queue was superior to the FIFO queue in all scenarios. In scenario 1, the rate of deadlines met was less than 20\% for both queues, so the MEC nodes performed an expressive number of forwardings when predicting that the deadlines would be missed. The preferred queue reduced the number of forwards with a difference of 2.61\%. In scenarios 2 and 3, there was a drastic reduction in the number of referrals. The preferred queue reduced the number of forwards with a difference of 6.49\% and 0.43\% for scenarios 2 and 3, respectively.

\subsection{Discussion}

The performance of the preferential queue was superior to that of the FIFO queue in terms of the number of referrals and the number of deadlines met. The superiority of the preferred queue is justified because it behaves like a FIFO queue in the worst case. That is, the requests that would have their deadline missed have the opportunity to be allocated in front of others already allocated. If the deadline of the other requests is affected, then the FIFO strategy is used to allocate the request at the end of the queue. This possibility of reallocation increases the number of deadlines met and reduces the number of referrals, as referrals are made only when there is a possibility of missing a deadline.
\section{Conclusion}
\label{sec:conclusion}

We explored the need for innovative strategies for distributed load orchestration through the optimization of the prioritization processing queue. We proposed a modification of a load distribution by introducing a loading orchestration mechanism based on preferential requisition queue. This strategy makes it possible to allocate new requisitions in front of those already allocated whenever the deadline for the others is not violated. 

We demonstrated that the performance of the proposed innovation was superior to that of the FIFO-based approaches in terms of the number of referrals and the rate of deadlines compliance. The results show that the preferential queue is more efficient than the FIFO queue, as it generated up to 6.49\% fewer referrals and had up to 5.97\% better deadlines compliance. The optimization is justified as the proposed innovation, in the worst case scenario when deadline-based prioritization is not viable, will behave as a FIFO-based queue.

In future work, we will explore other approaches for improving the algorithm around the \emph{prioritization processing queue}, such as the use of Big Data and machine-learning based models for prioritization recommendation. We also see a possibility to extend this approach beyond the problem of load orchestration, towards models for \emph{intrusion detection} \cite{vieira2019autonomic}. We intend to explore methods of Big Data and machine-learning to correlate thread signatures and workload distribution aiming to detect potential security threads.

\section{Acknowledgment}

This research was partially supported by a grant from the Coordination of Superior Level Staff Improvement (CAPES), Finance Code 001.

\bibliographystyle{IEEEtran}
\bibliography{main}

\end{document}